# The Rent's too High:
# Self-Archive for Fair Online Publication Costs

Robert T. Thibault*[a,b], Amanda MacPherson[a], Stevan Harnad[c,d], Amir Raz[a,b,e,f]

[a]McGill University, Montreal, QC, Canada

[b]Institute for Interdisciplinary Brain and Behavioral Sciences, Chapman University, Irvine, USA

[c]Département de psychologie, Université du Québec à Montréal, QC, Canada

[d]University of Southampton, UK

[e]Institute for Community and Family Psychiatry, Montreal, QC, Canada

[f]The Lady Davis Institute for Medical Research, Jewish General Hospital, Montreal, QC, Canada

*Please address correspondence to robert.thibault@mail.mcgill.ca

Draft dated: 17 August 2018

Even in the current climate of anti-science sentiment, science remains one of the most stunning achievements of our species. The main contributors of scientific knowledge—researchers—generally aim to disseminate their findings far and wide. And yet, publishing companies have largely kept these findings behind a paywall. With digital publication technology markedly reducing cost, this enduring wall seems disproportionate and unjustified.

The standard publishing model, *pay-to-access*, expects readers and their institutions to buy the articles they desire. This system provides the foundation for a commercial oligopoly—a small number of large sellers—to earn substantial profits from the work of scientists. Five companies publish over half of all scientific articles (*1*). Based on recent reports of annual profits, the biggest players—Elsevier, Wiley, and Springer Nature (merged since May 2015)—collected £913 million ($1274m), $687m, and nearly €600m ($714m), respectively (*2–4*)[†]. Who foots the bill? Academic institutions, largely. Approximately €7.6 ($8.3) billion goes into journal access every year (*5*). After paying these fees, even the wealthiest institutions gain access to only a fragment of the scientific literature; the less wealthy—a smidgeon.

Many of the services that publishing companies provide have fallen into obsolescence; however, one essential feature—arbitrating peer-review—remains key (*6*).  Even then, it's the researchers themselves who do the peer-reviewing free of charge.  Whereas subscription journals continue to charge exorbitant fees—perhaps because we agree to pay them—authors, institutions, and funding agencies seek mostly readership and impact, not financial gain.  In a word, publishing companies have co-opted our "give-away" research and left us with two major problems: accessibility and cost (*7*).

In an attempt to remedy the accessibility problem, an increasing proportion of journals now charge authors for publication, rather than readers for access.  This *pay-to-publish* model, known as "fee-based" gold open access (in contrast with "no-fee"—or more accurately named "subsidized"—gold open access, where journals publish freely available articles supported by alternate income sources) may increase accessibility; however, it only transfers the cost problem from libraries to authors, their institutions, and their funders—all of which rely heavily on public finances.  These gold publishers are no panacea; many charge from $2,500 to $5,000 per article (*8*) creating an environment ripe for "predatory" gold journals (offering "publishing" that amounts to no more than digital hosting with little or no quality control (*9*)).  The larger fee-based gold open access publishers, Public Library of Science (PLoS) and Frontiers, earned revenues of over $200m (*10*) and approximately $120m‡, respectively, in article processing charges over the past five years.  The open access arm of predominantly subscription-based publishers performed comparably: in the most recent year of record, Wiley earned $42m in revenues from open access fees (*3*); Elsevier $53.5m§.  Nature Publishing Group (NPG) garnered $54.5m‖ from its two flagship open access journals alone.  The revenues from fee-based gold open access continue to grow (see Figure 1).  These multi-million-dollar stakes engender a third problem: quality.

On the one hand, because fee-based gold open access publishers receive a payment for every article, a bias toward accepting rather than rejecting manuscripts may arise (*6*).  At its worst, this incentive lays the foundation for fraudsters to swindle unsuspecting researchers into publishing in predatory journals.  To be sure, when companies such as *Frontiers* publish pseudo-scientific claims of clairvoyance and propagate the vaccine-autism myth, the line between bona fide open access and predatory journals begins to blur.  On the other hand, gold open access does entail a degree of benefit.  If all *pay-to-access* journals adopted a *pay-to-publish* model, the accessibility problem would be solved, and, as long as funders continue to cover article processing charges, the cost problem would diminish to an extent (*11*). Pursuing this path, however, would only serve to delay the adoption of an open access publishing model at minimal cost for all parties. We would be building an infrastructure just to dismantle it shortly afterwards in the quest for an even better system. To counter this state of affairs, if enough authors made their publications publicly available by self-archiving their peer-reviewed drafts in institutional repositories at or before the date of official publication, the need for journal subscriptions and gold open access would quickly dwindle. Near-universal self-archiving would remove the paywalls associated with publishing and accessing academic articles, and in turn, establish research output as a public good. The market could then decide how much was worth paying for "fair" gold open access to cover the minimal costs of organizing and adjudicating peer-review. This change would unbundle access to the content of scientific articles, which requires little more than peer-review and a repository, from the superfluous aspects of publishing—including print copies, type-setting, marketing, and other expenses (such as CEO compensation ranging from just under $500,000 for PLOS (*12*) to $13.5m for Elsevier (*13*)). At the moment, gold open access appears to distract scientists with short-term improvements rather than near-optimal solutions. Widespread self-archiving would pave the road toward revamping our publishing system to a more equitable and sustainable state.

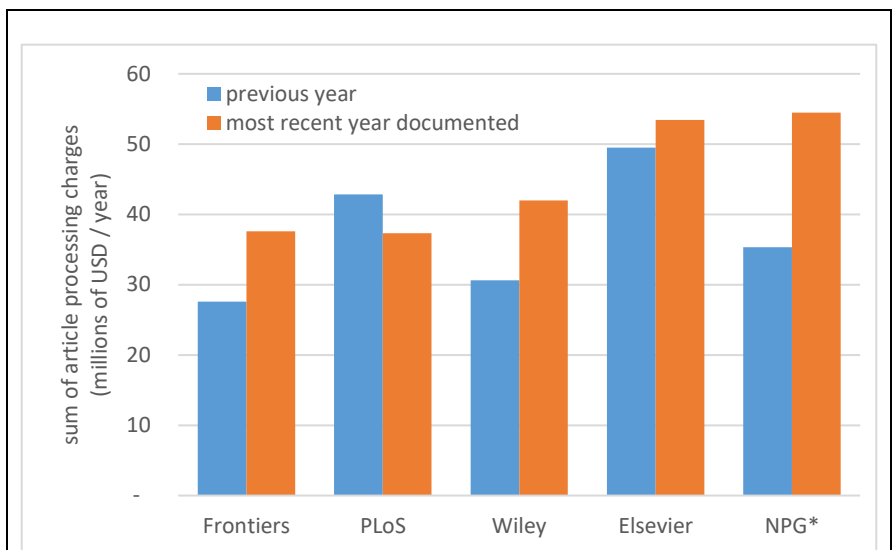

Figure 1. **The rising cost of fee-based gold open access.** In the past year, revenues from article processing charges from five leading companies raised 21%, from $186m to $225m.

\* These data represent revenues from the journals *Nature Communications* and *Scientific Reports* only.

Whereas both gold open access and self-archiving (often called "green" open access) boost the number of readers and citations an article receives (7), self-archiving can solve a fourth problem we have yet to discuss: delay. In standard academic publishing, a year or more can easily elapse from first submission to publication. With self-archiving, authors can upload their pre-refereeing preprint to a repository before they even submit it to a journal; and (in a spirit similar to that of gold open access journals that encourage post-publication peer-review) authors can continue posting updated versions as the review process advances. This publication model serves to benefit everyone.

The concept of self-archiving is far from new. Uploading manuscripts to openly accessible repositories began with the invention of the Web almost 30 years ago. For example, since 1991 physicists and mathematicians have been using *arXiv.org* to provide more than a million e-prints on a budget of just over $1m per year. Unfortunately, few researchers take this approach, even when their institutions mandate it, and most institutional repositories remain chronically underused. The few exceptions are repositories with effective mandates that generate high deposit rates (e.g., University of Liège and PubMed Central). These institutions and funders are leading the way by adopting and implementing verifiable mandates with incentive policies whereby only publications self-archived near the date of manuscript acceptance are eligible for institutional research performance evaluations for tenure and promotion or for eligibility to submit grant applications to funding agencies. Adopting these policies for self-archiving may solve many of our science-publishing concerns.

While some researchers oppose these measures, they often rely on faulty arguments concerning the economic and behavioral implications of mandatory self-archiving. The current status quo in scientific and scholarly journal publication is at odds with the idealized economic model in which, without regulation, a fair price emerges for almost any product (i.e., the free market system). For this system to work equitably, each party involved in creating a product must attempt to maximize their profit. In science publishing, however, neither authors nor reviewers ask for (or receive) financial compensation—they provide their services and expertise for the advancement of science and the benefit of humanity. Publishers, on the other hand, cash in on this scholarly product. The price tag on science publishing, moreover, conflates essential services and dispensable ones. Imagine entering a grocery store to buy food but finding that you only have the option to sit down and pay restaurant fees for service and preparation: a fair price cannot emerge for the groceries alone. Nor are mandates likely to deter researchers.

For example, if large funding agencies require self-archiving (and provide simple means to do so), researchers are unlikely to stop applying for their grants. Similarly, if a university requires providing green open access, the probability that academics will flee in search of a different home institution remains minimal. As a case in point, the institutions with the strongest and longest standing open access policies continue to thrive (e.g., University of Liège (*14*) and the Higher Education Research Funding Council for England [HEFCE] (*15*)).

Without stronger incentives for self-archiving, business will proceed as usual (*14*). Good will alone is unlikely to change publishing practices. Elsevier and Wiley maintain profit margins, of 37% (*2*) and 74% (*3*), respectively, that consistently outperform the most lucrative corporations, including Google and JP Morgan Chase. Like any company with a fiduciary duty to their shareholders, publishing giants need a market signal. If prominent funding agencies and leading research institutions provide mandates and compelling incentives for green open-access, for example by considering only manuscripts self-archived immediately upon acceptance in institutional research performance review and funder grant applications (*16*), the main product that subscription journals sell would markedly reduce in price (*17*). To cover the nominal cost necessary to assist and encourage self-archiving at their institution, libraries could, for example, cut a few journals from the $9m budget the average North American University spends on subscriptions each year (*18*).

Whereas science publishers have increasingly monetized academic research output since the 1950s, we are now at a crossroads. Springer Nature plans to join the corporate ranks of Elsevier and Wiley with an Initial Public Offering (even if delayed by market conditions); the Dutch government recently locked in to a fee-based gold open access deal with major publishers wherein they pay €1300-2000 ($1500-2400) from public funds for each article their researchers publish, and Germany has shown reluctance to follow suit (*19*). As this new infrastructure takes shape, we must note: many for-profit corporations have done little more than repackage the sale of our scholarly product from subscription premiums to article processing charges—they continue to reap hefty profits riding on the coattails of idealistic (or under-informed) scientists. While the new fee-based golden open access wrapping does entail some benefits, widespread self-archiving can more effectively return research output to the hands of scientists. It can reduce cost until a reasonable price emerges for post-green gold open access, thereby promoting quality, and minimizing delay and further increasing accessibility,

The history of open access reveals a disheartening irony. Physicists invented the World Wide Web to share their research efficiently; the sluggish workflow and static output of the printing press hindered their progress. Nearly three decades later, we use the Web for everything from *a-to-z* but have yet to realize the full potential of its original purpose: to share research output swiftly and cheaply. Instead, we pay exorbitant fees to access far too few research findings. That the "rent is too damn high" (cf. Jimmy McMillan) should be plain to see; stronger self-archiving policies that would deflate current science publishing costs should be easy to put into practice.

**Footnotes:**

[†]The figures come from the *Science, Technical, and Medical* subdivision of the RELX group annual report and the *Research* subdivision of Wiley's annual report. Profits are calculated as total revenues minus total costs. Profit margins are calculated as profits divided by total revenue. We performed all currency conversions based on the historic exchange rate (according to *xe.com*) on the date each figure was published.

[‡]To calculate this sum we multiplied the article processing charges available from the Frontiers website (taken on 30 April 2017) by the number of articles they published (accounting for the different prices based on article type and journal). Based on the discounts we could find for 2014 ($1.9m), 2016 ($3.1m), and 2017 ($5.0m) we assumed a similar 10% discount from the total sum for 2013 and 2015.

[§]To calculate this sum we multiplied numbers taken from Elsevier's annual financial report and website: 27,000 open access articles in 2017 by an average article processing charge of $1,980 (G. Hersch, Facts dispel false price point reported by Science Magazine. *Elsevier*, 2017).

[ǁ]This sum represents the number of citable items in *Nature Communications* and *Scientific Reports* for 2016 taken from Thompson Reuters InCites Journal Citation Reports multiplied by the article processing charges of $5,200 and $1,760, respectively. Springer Nature publishes an additional 208 (Springer) and 48 (NPG) open access journals that were not included in this figure.

**Acknowledgements:**

We thank Peter Dain Suber, Stuart M. Shieber, and Michael Ira Posner for constructive comments on earlier drafts of this manuscript.